\documentclass[pre,twocolumn,superscriptaddress,showpacs]{revtex4}

\usepackage[dvips]{graphicx}
\usepackage{amsmath}

\def\0{\varnothing}

\begin{document}

\title{Critical phase in non-conserving zero-range processes and equilibrium networks}
\author{A.G. Angel}\author{M.R. Evans}
\affiliation{School of Physics, University of Edinburgh, Mayfield
Road, Edinburgh EH9 3JZ, UK} \author{E.
Levine}\affiliation{Department of Physics of Complex Systems, Weizmann Institute of Science, Rehovot, Israel 76100}\affiliation{Center for Theoretical Biological Physics,
University of California at San-Diego, La Jolla, CA 92093
}\author{D. Mukamel}
\affiliation{Department of Physics of Complex Systems, Weizmann Institute of Science, Rehovot, Israel 76100}

\begin{abstract}
Zero-range processes, in which particles hop between
sites on a lattice,
are closely related to equilibrium networks, in which
rewiring of links take place. Both systems exhibit a condensation
transition for appropriate choices of the dynamical rules. The
transition results in a macroscopically occupied site for zero-range processes
and a
macroscopically connected node for networks. Criticality, characterized by a
scale-free distribution, is obtained only at the transition point.
This is in contrast with the widespread scale-free real-life
networks. Here we propose a generalization of these models whereby
criticality is obtained throughout an entire phase, and the
scale-free distribution does not depend on any fine-tuned
parameter.
\end{abstract}

\pacs{89.75.-k, 05.70.Ln, 05.40.-a}

\maketitle

Many driven, non-equilibrium models, reach a critical or scale
invariant steady state only when their dynamical parameters are
fine-tuned to reach a phase transition point. Examples 
include a
wide range of systems such as jamming in traffic \cite{OEC98},
coalescence in granular gases \cite{E99}, gelation in networks
\cite{KRL00} and wealth condensation in macroeconomics
\cite{BJJKNPZ02}.
In all these systems one has a condensation phase transition
which we shall discuss in detail below.
In other non-equilibrium models, for example in
driven lattice gases and sandpile models \cite{SZ,SOC}, it has
been argued that scale invariance and power-law distributions are
generic, or at least one may have scale-free distributions across
wide regions of the parameter space rather than just at critical
points. This phenomenon has been termed self organized criticality.

In recent years considerable attention has been given to the study
of real-life networks. Networks, defined as  collections of nodes
connected by links, are found in many fields of study, ranging
from molecular biology to social communities and the Internet
\cite{NETS,DM03}. With each node one associates a degree $k$ which
is the number of links connected to it. In general links may be
directed or they may carry a weight, however for our purposes we
do not consider such features. It has been observed that very
often real-life networks are characterized by a degree
distribution $p(k)$ which decays algebraically for large $k$
\cite{NETS,DM03}. These networks, termed scale-free networks, are
indeed critical, suggesting the existence of a mechanism  which
drives them to this state. Subsequently, dynamical processes for
{\it growing} networks have been proposed which result in a
critical distribution for a wide range of the dynamical
parameters. In these processes nodes and links are continually
added to the network with some predetermined rates
\cite{KRL00,AB,KR01,BB01}. On the other hand ``equilibrium"
networks \cite{DMS03,DM03}, whose dynamics constitutes rewiring
processes with a fixed number of nodes, exhibit a critical
distribution only at a critical point. This transition corresponds
to condensation (also referred to as gelation) where a single node
captures a finite fraction of the links. Note that although termed
equilibrium networks, their dynamics does not always obey detailed
balance, thus their steady state may not always be a thermal
equilibrium one.

Instructive insight has been gained into the condensation transition
through the analysis of simple interacting particle systems
\cite{EH05,BBJ,EMZ05}.  These systems form fundamental models which
may be mapped onto particular applications.  For example, the
zero-range process (ZRP) \cite{E00} is a particularly simple and
exactly soluble model in which each site $\mu$ of a lattice may
contain an integer number of particles $n_\mu$ and particles hop to a
neighboring site with rate $u(n_\mu)$. This model is closely related
to the equilibrium networks discussed above which undergo rewiring
dynamics \cite{DMS03}; in the following we will exploit this
relationship.

Condensation in the ZRP will occur, for example, when $u(n)$ decays to
some finite, large-$n$ asymptotic value $\beta$, as $u(n) \sim
\beta(1+b/n)$ with $b>2$.  The transition is simply understood by
considering $p(n)$ the steady-state probability that a site contains
$n$ particles.  For a subcritical density of particles one finds
that $p(n)$ decays exponentially with decay length dependent on the
conserved particle density $\rho= N/L$ where $N$ is the number of
particles and $L$ is the number of sites.  This $p(n)$ describes the
low density, fluid phase. As the density is increased the decay length
increases until at the critical density, $\rho_c$, it diverges and one
has a power-law distribution $p(n)\sim n^{-b}$, thus a critical fluid.
Above $\rho_c$, in addition to the power law, a piece of $p(n)$
emerges centered about $n=L(\rho-\rho_c)$; this piece represents the
condensate \cite{EMZ05}. Thus the supercritical phase corresponds to a
critical fluid coexisting with a condensate. Only at criticality does
one have a pure power-law distribution.

In the present work we investigate how a critical {\em phase} may emerge
in processes such as ZRPs and network models. We shall see that
the introduction of non-conservation in an appropriate
fashion, modifies the condensate phase into a scale-free phase by
effectively suppressing the condensate and leaving the critical
fluid. In this generalization of the ZRP particles are created at
all sites at constant rate but are removed at a rate that depends
non-linearly on the occupation number. Equivalently in the network
context links are created and destroyed. We begin by  elucidating
the mechanism for the generation of a critical phase within a
generalized  ZRP; later we will define an equilibrium network
where this mechanism is also manifested.

Consider a lattice of $L$ sites upon which reside a number of
particles. With rate $u(n_\mu)$ (probability per unit time) which
depends on the occupation $n_\mu$ of  site $\mu$, a particle is
transferred from site $\mu$  to another site. We consider a fully
connected geometry, where the destination site is  chosen randomly
from the other $L-1$ sites. In addition to the hopping dynamics, 
particles are added to site $\mu$ with a constant rate $c$, or
removed with a rate  $a(n_\mu)$, which increases with the site
occupation  $n_\mu$. Thus, our choice for the dynamical rates is
given by
\begin{eqnarray}
n, m \to n-1, m+1\;&\mbox{with rate}&\;
\mbox{$u(n) = \left(1 + \frac{b}{n}\right)\theta(n)$} \nonumber \\
n \to n+1 \;&\mbox{with rate}&\;  \mbox{$c = \left(\frac{1}{L}\right)^{s}$}
\label{dyn}
\\
n \to n-1 \;&\mbox{with rate}&\;
\mbox{$a(n) = \left( \frac{n}{L} \right)^{k}$}\;, \nonumber
\end{eqnarray}
where $b$, $s$ and $k$ are positive parameters, and $\theta(n)$ is the
usual Heaviside step function. The dynamical rates are conveniently
implemented by using a random-sequential updating scheme, whereby
at each time step a site $\mu$ is chosen at random, and a hop,
annihilation or creation event may occur with relative
probabilities given by the rates (\ref{dyn}). With $b\leq2$ the
system is always found in a sub-critical phase, where the particle
number distribution is exponential. We therefore restrict our
discussion hereafter to the case $b>2$.

In Fig.~1 we compare the particle number distribution $p(n)$ of
our model with that of the ZRP with conserving dynamics. It is
clearly seen that for this choice of parameters, the
creation/annihilation dynamics can selectively destroy any
condensate and sustain the power-law distribution, corresponding
to the critical fluid. In what follows we analyze the model
showing that this feature holds for an entire region in the
parameter space.

\begin{figure}[t]
\label{nocacomp}
\begin{center}
\includegraphics[width=7cm]{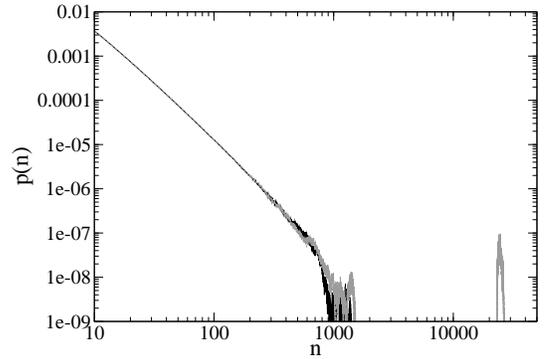}
\caption{Steady state distribution of the non-conserving ZRP
  (black solid line) and  the conserving ZRP
  (grey dotted line).  The data is from simulations run on a system of
  size $L=10^4$, with $b=2.6$, $k=3$ and $s=1.96$.  In the conserving
  model the particle density was set to $\rho=4> \rho_c$.  The peak at
  high occupation number, which exists only in the conserving model,
  corresponds to the condensate.}
\end{center}
\end{figure}

The steady state of the model is fully described by the probability
distribution $P(n_1,n_2,\ldots,n_L)$ over all possible configurations.
In contrast to the conserving ZRP \cite{E00}, the steady-state
distribution of the model (\ref{dyn}) does not factorize
generally. However, we make the mean field approximation that the
steady state distribution does factorize, i.e.
$P(n_1,n_2,\ldots,n_L) \to  \prod_{i=1}^L p(n_i)$.
Due to the fully-connected geometry
we expect this  approximation to become  exact in the limit $L\to \infty$.

Using this approximation, the steady-state master equation is given
by
\begin{eqnarray}
0 &=& \left[ u(n+1) + a(n+1) \right]p(n+1)
- \left[ \lambda + c \right] p(n) \label{mastereqn1}\\
&&- \left\{ \left[ u(n) + a(n) \right]p(n) - \left[\lambda +
c\right]p(n-1)\right\} \theta(n).  \nonumber 
\end{eqnarray}
Here the current $\lambda$ is given by
\begin{equation}
\lambda = \sum_{n=1}^{\infty} u(n) p(n)\;. \label{lambdaeqn}
\end{equation}
From (\ref{mastereqn1}) it follows that
\begin{equation}
p(n) = \frac{\left(\lambda+c\right)^n} {\prod_{m=1}^n \left[ a(m)
+ u(m) \right]}\;p(0)\;. \label{ssP}
\end{equation}
Note that this is not a closed solution as $\lambda$ depends on
$p(n)$. The values of $\lambda$ and $p(0)$ should be set such that
both the normalization condition
$1 = \sum_{n=0}^{\infty} p(n)$
and the creation/annihilation balance condition
\begin{equation}
 c = \sum_{n=1}^{\infty} a(n) p(n)
\label{genbalanceeqn}
\end{equation}
are obeyed.
We now identify the three  phases of the
model by determining the asymptotic, large $L$ behaviors of $p(n)$
and $\lambda$ which satisfy (\ref{lambdaeqn}) and (\ref{genbalanceeqn}).
The emergent phase diagram is summarised in Fig.~2.
Deferring details to a later publication,
we find the following results:

{\bf Low-density phase, $s>k$} --- Rewriting \eqref{genbalanceeqn}
as $L^{k-s}=\sum{n^kp(n)}$ implies
$p(n)$ is a rapidly decreasing
function of $n$, and the steady state density $\rho$ is $\ll
1$. Thus, $p(1) \simeq \rho \sim L^{k-s}$ and in the
thermodynamic limit the density goes to zero.

For $s<k$ the sum in
\eqref{genbalanceeqn}  is controlled by the behavior of $p(n)$ at
large $n$. We find the following regimes:

{\bf High-density phase, $s<k/(k+1)$} ---
Here
\begin{equation}
p(n) \sim
n^{-b} \exp \left[ \frac{n}{L^s}
 -\frac{n^{k+1}}{(k+1)L^k} \right]\;,
\end{equation}
thus
$p(n)$ is strongly peaked at $n \sim L^{1-s/k}$. This is
the high density phase, where all sites are highly occupied. Note
that the mean number of particles in the system, $N \sim
L^{2-s/k}$ is super-extensive.

{\bf Critical phase, $k > s > k/(k+1)$} --- In this phase the system
relaxes to the critical density, and $p(n)$ takes an algebraic form.
However, for large but finite systems two sub-phases are observed,
distinguished by the finite-size corrections to the dominant power law.

For $k>s>kb/(k+1)$
\begin{equation}
p(n) \sim
n^{-b} \exp \left[ -\frac{n}{L^x} \right]\;,
\end{equation}
where $x = (k-s)/(k-b+1)$. This
cuts off the power law at $n\sim L^x$.
We refer to this as  critical
sub-phase (a).

For $kb/(k+1)>s>k/(k+1)$
\begin{equation}
p(n) \sim
n^{-b} \exp \left[ n d \left(\frac{\ln L}{L} \right)^{\frac{k}{k+1}}
 -\frac{n^{k+1}}{(k+1)L^k} \right]\;,
\end{equation}
where $d = b-s(k+1)/k$.
Here,
on top of the algebraic part, $p(n)$ is weakly peaked at $n\sim
L^{k/(k+1)}(\ln L)^{1/(k+1)}$. This peak will diminish as $L \to
\infty$. We refer to this as
critical sub-phase (b).

\begin{figure}[t]
\begin{center}
\includegraphics[width=5cm]{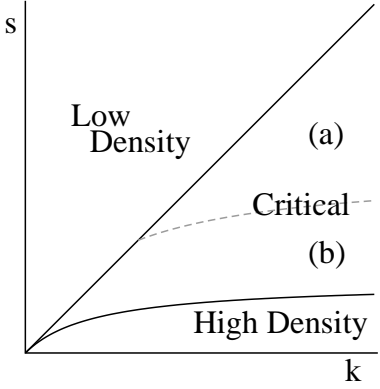}
\caption{Typical phase diagram of both
ZRP and network models in the
$k$-$s$ plane
with $b$ fixed,
$k$,$s$,$b$ are  defined in (\ref{dyn}).}
\end{center}
\label{phasediag}
\end{figure}
In Fig.~3 we present typical data
obtained from numerical
simulations in the different phases and compare with
theoretical curves of $p(n)$.
We found that in all
phases, starting from random initial configurations of various
densities, the system relaxes towards its expected steady
state value. However, for the low-density phase the time scales
for full relaxation were prohibitive and we do not present
steady state data for this phase.  In Fig.~3 we also provide  data for a
one-dimensional (1d) variant of the model, where sites are
arranged in a 1d array, and particles are allowed to
hop only to the right neighbor of the departure site.
For the 1d geometry the mean-field
approximation is not expected to be exact even
in the limit $L \to \infty$.
Nevertheless, we find numerically that the three  phases
discussed above exist also in the 1d model.

\begin{figure}[t]
\begin{center}
\includegraphics[width=8.5cm]{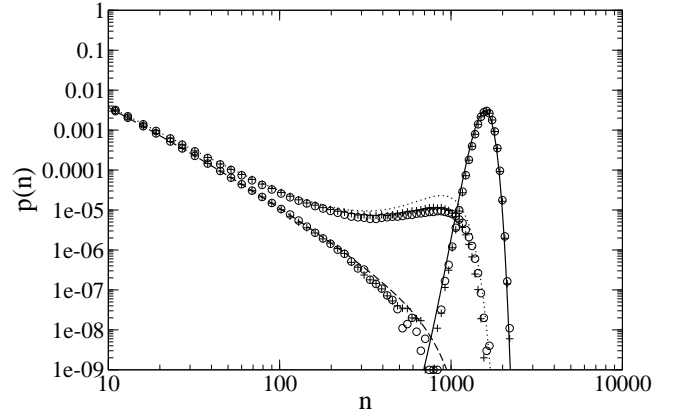}
\caption{Steady state  distributions from simulations of the ZRP
model on a fully-connected lattice (+) and a 1d lattice (O),
compared with the theoretical asymptotic curves.  Here $L=5000$
and $b=2.6$, $k=3$.  Dashed curve is critical sub-phase (a)
($s=2$); dotted curve is critical sub-phase (b) ($s=1.2$); full
line is high-density phase ($s=0.4$).}
\end{center}
\label{numphases}
\end{figure}

We now apply the approach discussed above to equilibrium networks.  To
make the analogy with the ZRP, one identifies a site of the ZRP and
its occupation number with a node in the network and its degree,
respectively.  We define a network model which incorporates both
rewiring and creation/annihilation dynamics, and show how a proper
choice of rates leads to the existence of a critical phase, much like
that of the non-conserving ZRP, within which networks are scale free.
Due to the introduction of annihilation of links, there is no simple
mapping from the  ZRP to the network model since this would require
keeping track of pairs of linked particles in the ZRP. 
However the two systems 
are closely related and as we shall see share the same phase diagram.

We consider a network of
$L$ nodes which are linked together by an integer number, $N/2$,
of undirected links ($N$ is the number of particles in the
corresponding ZRP).  With rate $u(n_\mu)=1+b/n_\mu$
one of the $n_\mu$ links is disconnected from node $\mu$ and is rewired
to another randomly chosen node. This does not change the number
of links in the network. With rate $a(n_\mu)=(n_\mu/L)^k$ one of the links
connected to node $\mu$ is removed from the network. In addition,  a new link
is created between node $\mu$ and other randomly chosen node
at a constant rate $c=1/L^s$.
Again the dynamics is conveniently implemented by
choosing a node $\mu$ randomly at each time step,
and  changing the wiring with probabilities constructed from the relevant
rates.
\begin{figure}[t]
\begin{center}
\includegraphics[width=8.5cm]{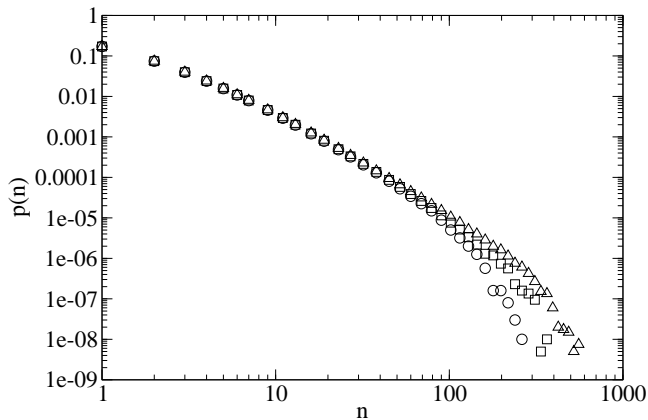}
\caption{Steady state probability distributions from simulations
of the network model in the critical phase (sub-phase (a)).
Here $b=2.6$, $k=3$ and $s=2$,
with $L=1000$ (circles), $L=2000$ (squares) and $L=4000$ (triangles).}
\end{center}
\label{networkdata}
\end{figure}
The mean-field master equation for the network model differs slightly from
that of the ZRP \eqref{mastereqn1}, and is given by
\begin{eqnarray}\label{netmastereqn}
0 &=& \left[a(n+1)+u(n+1)+
\Lambda(n+1)\right] p(n+1) \nonumber \\
&-& \left[a(n)+u(n)+\Lambda(n)\right]p(n)\theta(n) \\
&-& \left[\lambda + 2c\right]p(n)  + \left[\lambda +
2c\right]p(n-1)\theta(n) \nonumber\;.
\end{eqnarray}
Here $\lambda = \sum u(n)p(n)$ as before, and
\begin{equation}
\Lambda(n) = \frac nN \sum_{l=1}^{\infty} a(l)p(l) = c\frac nN\;.
\end{equation}
The steady state solution to (\ref{netmastereqn})  is given by
\begin{equation}
p(n) = \frac{(\lambda + 2c)^n}{\prod_{m=1}^n \left[\Lambda(m) + a(m) +
u(m)\right]} p(0)\;.
\end{equation}
The main difference between this result and that of the ZRP
(\ref{ssP}) lies in  $\Lambda(n)$
and its  dependence on the total
number of links in the system. This complicates slightly the analysis,
however all phases persist including the critical phase.
Thus the phase diagram in Fig.~2 also describes the network model:
the low-density and high-density phases
characterize extremely sparse and extremely dense networks; the critical fluid
corresponds to  a scale-free phase in the thermodynamic limit.
Examples of the degree
distribution in the critical phase are given in Fig.~4, where we present
data obtained from  simulations of systems of increasing
sizes. The power-law regime increases with system size.

Further interesting observations are that the weak peak of $p(m)$ in
critical sub-phase (b) may correspond to a number of highly connected
nodes but is distinct from a condensate phase.  This  would correspond
to a large number of hubs
in the network.  Also, we note that in the network model
the suppression of events which link nodes to themselves has a
considerable effect on the results \cite{BPV}, by introducing a new
cut-off into the probability distribution which can become the
dominant scale. A full analysis will be published elsewhere.

Our main interest in the systems studied lies in the emergence of a
critical phase which we have shown exists for annihilation and
creation indices $k$, $s$ in the range $k>s> k/(k+1)$. We conclude by
comparing the critical phases we have identified to the critical
points of the corresponding conserving ZRP and network models. In the
latter, the average particle/link density $\rho$ is an external
parameter. A power-law distribution of the occupation number/degree is
only obtained at $\rho=\rho_c$. In contrast, in the non-conserving
models we have studied, the steady-state density is set by the
dynamics to be $\rho_c$ and a power-law distribution is obtained
throughout the critical phase. In models exhibiting self-organized
criticality \cite{BDMZ}, a critical phase is typically obtained
only when the driving rate of the system vanishes with the system size
\cite{SJD},
in order to ensure relaxation between stimuli.
In comparison, in the present work the
creation rate $c$  in e.g.\ (\ref{dyn}), which  could be thought of as a
driving rate for the system,  vanishes  in the large $L$ limit
whereas $u$ the hopping/rewiring rate does not vanish.
Thus, there is a separation of timescales in the dynamical processes.
On the other hand, there are no avalanches or underlying absorbing states
which are features usually associated with self organized criticality.

This study was supported by the Israel Science Foundation (ISF).
Visits of MRE to the Weizmann Institute were supported by the
Albert Einstein Minerva Center for Theoretical Physics. A visit
of DM to Edinburgh was supported by EPSRC programme grant GR/S10377/01.

%


\begin{thebibliography}{999}
\bibitem{OEC98}
O. J.~O'Loan, M. R.~Evans and M. E.~Cates {\it Phys. Rev. E} {\bf 58}, 1404 (1998).

\bibitem{E99}
J. Eggers {\it Phys. Rev. Lett.} {\bf 83}, 5322 (1999).

\bibitem{KRL00}
P. L. Krapivsky, S. Redner and F. Leyvraz
Phys. Rev. Lett. {\bf 85}, 4629 (2000)

\bibitem{BJJKNPZ02}
Z. Burda {\em et al}
(2002)  {\it Phys. Rev. E} {\bf 65}, 026102

\bibitem{SZ}
B Schmittmann and R K P Zia 1995 {\it Statistical Mechanics of
Driven Diffusive Systems} ed C Domb and J L Lebowitz (Academic
Press, New York)

\bibitem{SOC} P. Bak, C. Tang, and K. Wiesenfeld, {\it Phys. Rev. Lett.} {\bf59}, 381 (1987);
P. Bak, {\it How Nature Works} (Copernicus press, NY, 1996);
H.J. Jensen, {\it Self-Organised Criticality} (CUP, 1998)

\bibitem{NETS} For recent reviews see R. Albert and A.-L.
Barab\'asi, {\it Rev. Mod. Phys.} {\bf74}, 47 (2002); M.E.J.
Newman, {\it SIAM Review} {\bf 45}, 167 (2003).


\bibitem{DM03}
S.N. Dorogovtsev and J.F.F. Mendes, {\it Evolution of Networks},
(Oxford, OUP, 2003)

\bibitem{AB}
A.-L. Barab\'asi and R. Albert, {\it Science} {\bf 286}, 509
(1999).

\bibitem{KR01}
P. L. Krapivsky and S. Redner Phys. Rev. E {\bf 63}, 066123 (2001)

\bibitem{BB01}
G. Bianconi and A-L Barab\'asi Phys. Rev. Lett. {\bf 86}, 5632
(2001)

\bibitem{DMS03} S.N. Dorogovtsev, J.F.F. Mendes, A.N. Samukhin, {\it Nucl.
Phys. B} {\bf 666}, 396 (2003).



\bibitem{EH05}
{\it For a recent review see}
M. R. Evans and T. Hanney (2005) Cond-mat/0501338


\bibitem{BBJ}  P. Bialas, Z. Burda, and D. Johnston, Nucl. Phys. B {\bf
493}, 505 (1997).

\bibitem{EMZ05}
S.N. Majumdar, M. R. Evans,  R. K. P. Zia
Cond-mat/0501055

\bibitem{E00}
M. R.~Evans {\it Braz. J. Phys.} {\bf 30}, 42 (2000).



\bibitem{BDMZ}
A. Vespigniani, R. Dickman,M. A. Mu÷noz, and S. Zapperi, {\it Phys. Rev. Lett.} {\bf 81}, 5676 (1998)

\bibitem{SJD}
D. Sornette, A. Johansen and I. Dornic, {\it J. Phys. I (France)} {\bf 5}, 325 (1995).


\bibitem{BPV}
M. Boguna, R. Pastor-Satorras and A. Vespignani,
{\it Euro. Phys. J. B} {\bf 38}, 205 (2004)
\end{thebibliography}
\end{document}